\RequirePackage{ifpdf}
\ifpdf 
\documentclass[pdftex]{sigma}
\else
\documentclass{sigma}
\fi

\usepackage{amsbsy,eucal,mathrsfs}

\newcommand{\qu}{\mbox{\small$\frac{1}{4}$}}
\newcommand{\se}{\mbox{\small$\frac{3}{2}$}}
\newcommand{\im}{{\rm i}}

\newcommand{\pa}{\shortparallel}
\newcommand{\ort}{\perp}
\newcommand{\cc}[1]{\lefteqn{\stackrel{*}{#1}}\phantom{#1}}
\newcommand{\lab}[1]{\label{#1}}
\newcommand{\re}[1]{(\ref{#1})}
\newcommand{\nn}{\nonumber}
\newcommand{\B}[1]{\boldsymbol{#1}}
\newcommand{\s}[1]{{\textbf{\sffamily{#1}}}}
\newcommand{\of}{\,\hat{\mathsf o}}
\newcommand{\ef}{\,\hat{\mathsf e}}
\newcommand{\sL}{\textbf{\sffamily{\symbol{"03}}}}
\newcommand{\sFi}{{\textbf{\sffamily{\symbol{"08}}}}}
\newcommand{\sPsi}{{\textbf{\sffamily{\symbol{"09}}}}}
\newcommand{\sD}{{\textbf{\sffamily{\symbol{"01}}}}}
\newcommand{\sO}{{\textbf{\sffamily{\symbol{"0A}}}}}
\newcommand{\cL}{{\mathsf\Lambda}}
\newcommand{\cS}{{\mathsf\Sigma}}
\newcommand{\cFi}{{\mathsf\Phi}}
\newcommand{\cPsi}{{\mathsf\Psi}}
\newcommand{\cf}[1]{\mathsf{#1}}
\newcommand{\iPhi}{\mathit\Phi}
\newcommand{\iPsi}{\mathit\Psi}
\newcommand{\sY}{{\mathscr Y}}
\newcommand{\mcl}[1]{{\mathcal#1}}

\begin{document}

\renewcommand{\textfraction}{0.01}
\renewcommand{\topfraction}{0.99}


\renewcommand{\PaperNumber}{048}

\FirstPageHeading

\renewcommand{\thefootnote}{$\star$}

\ShortArticleName{Solvable Two-Body Dirac Equation}

\ArticleName{Solvable Two-Body Dirac Equation\\ as a Potential
Model of Light Mesons\footnote{This paper is a contribution to the
Proceedings of the Seventh International Conference ``Symmetry in
Nonlinear Mathematical Physics'' (June 24--30, 2007, Kyiv,
Ukraine). The full collection is available at
\href{http://www.emis.de/journals/SIGMA/symmetry2007.html}{http://www.emis.de/journals/SIGMA/symmetry2007.html}}}

\Author{Askold DUVIRYAK}

\AuthorNameForHeading{A. Duviryak}

\Address{Institute~for~Condensed~Matter~Physics of National
Academy of Sciences of Ukraine, \\ 1 Svientsitskii~Str., UA--79011
Lviv, Ukraine}

\Email{\href{mailto:duviryak@ph.icmp.lviv.ua}{duviryak@ph.icmp.lviv.ua}}

\ArticleDates{Received October 29, 2007, in f\/inal form May 07,
2008; Published online May 30, 2008}

\Abstract{The two-body Dirac equation with general local potential
is reduced to the pair of ordinary second-order dif\/ferential
equations for radial components of a wave function. The class of
linear + Coulomb potentials with complicated spin-angular
structure is found, for which the equation is exactly solvable. On
this ground a relativistic potential model of light mesons is
constructed and the mass spectrum is calculated. It is compared
with experimental data.}


\Classification{81Q05; 34A05}

\renewcommand{\thefootnote}{\arabic{footnote}}
\setcounter{footnote}{0}

\section{Introduction}

    It is well known that light meson spectra are structured into Regge
trajectories which are approximately linear and degenerated due to
a weak dependence of masses of resonances on their spin state.
These features are reproduced well within the exactly solvable
simple relativistic oscillator model (SROM)
\cite{K-N73,Tak79,I-O94} describing two Klein--Gordon particles
harmonically bound.

    Actually, mesons consist of quarks, i.e., spin-1/2 constituents.
Thus potential models based on the Dirac equation are expected to
be more appropriate for mesons spectroscopy. In recent years the
two body Dirac equations (2BDE) in dif\/ferent
formulations\footnote{Here we mention three formulations of 2BDE
approach. One of them \cite{B-K85,B-U87,Gra91,Duv05} is built as a
generalization of the Breit equation \cite{Bre29,B-S57}, two
other, \cite{Saz86,M-S94,M-S95} and \cite{KvA87,KvA90,KvA99},
originate from Dirac constraint theory.} with various conf\/ining
potentials are used as a relativistic constituent quark models
\cite{Kro76,M-B83,KvA84,KvA87a,Chi87,Bra87,CLS91,SCS93,S-C93,Tsi00,M-N05}.
Some models are universal \cite{KvA04,Duv06}, i.e., describe
several states of heavy as well as light mesons. The solution of
2BDE is usually obtained by means of perturbative or numerical
methods, even in the case of simple potentials. As far as exactly
solvable models are concerned, only few examples are known in the
literature which represent versions of two-body Dirac oscillator
\cite{Saz86a,Saz88,MLV91,M-M94,MQS95}. Similarly to SROM, they all
exhibit Regge trajectories which are linear exactly or
asymptotically (for large values of angular momentum). But energy
levels are spin-dependent, and a type of degeneracy is dif\/ferent
from that of SROM and experimental data.

    In the present paper we attempt to construct, on the base of 2BDE,
the consistent solvable potential model which describes as well as
possible actual spectra of light mesons. For this purpose we start
from 2BDE with general two-fermion local potential proposed by
Nikitin and Fushchich \cite{N-F91}. Following the scheme
\cite{D-D02,Duv05,Duv06}, we perform the radial reduction of
general 2BDE and get a set of 8 f\/irst-order ODEs, then 4 of
f\/irst-order ODEs and f\/inally a pair of second-order ODEs. This
chain of transformations permit us to write down the set of
equations in a compact form and then to impose constraints for
general potential which considerably simplify the equations.

    In such a way we f\/ind a family of exactly solvable models with
linear potentials, which includes two known examples of the Dirac
oscillators \cite{Saz86a,MLV91,M-M94,MQS95} and new ones. Then we
construct integrable extensions of these models describing linear
plus Coulomb-like interaction of constituents and giving
asymptotically linear Regge trajectories. Free parameters can be
used to provide a spin-independent degeneracy for asymptotics of
trajectories. It is remarkable that one of the models reproduces
exactly the spectrum of SROM. It describes two equal-mass Dirac
particles bound by rather non-trivial potential.

    Calculated spectra are compared to experimental data for
the family of ($\pi$-$\rho$)-mesons.

    The Nikitin--Fushchich anzatz as well as our models are rotary invariant
but not Poincar\'e-invariant, as expected of consistent
description of relativistic systems. Thus we propose in Appendix
the manifestly covariant equations reduction of which into the
center-of-mass reference frame restores the former general
equation with the residual rotary symmetry. In this form our
models are convenient to compare (and some of them are found
equivalent) to other covariant two-body Dirac models known in a
literature.

\section{General structure of a potential}

    We start with the standard three-dimensional formulation of the two-body Dirac
equation. It has the form:
\begin{gather}\lab{2.1}
\left\{h_1(\B p) + h_2(-\B p) + U(\B r) - E\right\}\iPhi(\B r) =
0,
\end{gather}
where $\iPhi(\B r)$ is 16-component wave function (Dirac
4$\times$4-bispinor) of relative position vector $\B r$,
\begin{gather}\lab{2.2}
h_a(\B p) =\B\alpha_a\cdot\B p + m_a\beta_a
\equiv{}-\im\B\alpha_a\cdot\B\nabla + m_a\beta_a, \qquad a=1,2,
\end{gather}
are Dirac Hamiltonians of free fermions of mass $m_a$, $\B
p=-\im\B\nabla$, and $U(\B r)$ is an interaction potential. If
$\iPhi(\B r)$ is presented in $4\times4$-matrix representation,
the operators $\B\alpha_a$ and $\beta_a$ act as follows:
$\B\alpha_1\iPhi=\B\alpha\iPhi$,
$\B\alpha_2\iPhi=\iPhi\B\alpha^{\rm T}$ etc, where $\B\alpha$ and
$\beta$ are Dirac matrices.

    The potential $U(\B r)$ is a multiplication operator in the position
representation, which is invariant under spational rotations and
inversion. The most general form of such a potential is found in
\cite{N-F91}
\begin{gather}\lab{2.3}
U(\B r)=\sum_{A=1}^{48} U_A(r)\Gamma_A.
\end{gather}
It is parameterized by 48 arbitrary (complex, in general)
functions $U_A(r)$ of $r=|\B r|$ (we will refer to them as {\em
partial potentials}), and matrices $\Gamma_A$ are built in terms
of Dirac matrices and unit vector $\B n=\B r/r$.

    We require of the potential $U$ to be Hermitian with respect to the inner product
\begin{gather}\lab{2.4}
\langle\iPsi|\iPhi\rangle=\int d^3r\, \mathrm{Tr}(\iPsi^\dag(\B
r)\iPhi(\B r)).
\end{gather}
Consequently, the equation \re{2.1} becomes Hamiltonian. This
requirement reduces an arbitrariness in the general potential
\re{2.3} to 48 real partial potentials.

    Let us construct a Hermitian basis for matrices $\Gamma$ involved in the potential \re{2.3}.
For this purpose we start from the relation for Dirac matrices
$\B\alpha=\gamma^5\B\sigma$, where components $\sigma_i$
($i=1,2,3$) of the vector $\B\sigma$ are understood as either
Pauli 2$\times$2-matrices or block-diagonal 4$\times$4-matrices
$\mathrm{diag}(\sigma_i,\sigma_i)$. By this $\B\alpha$ is split
into two factors, one of which acts on ``particle-antiparticle''
degrees of freedom while another one -- on spin degrees of
freedom. We will refer to them as {\em Dirac} and {\em spin}
factors respectively. Similar splitting can be done for arbitrary
Dirac operator. Now, taking into account that $\beta$ is scalar,
$\gamma^5$ is pseudo-scalar, $\B n$ is vector and $\B\sigma$ is
pseudo-vector with respect to $O(3)$ group, we write down the
basis for $\Gamma$-matrices in the product form
\begin{alignat}{3}
& \Gamma_{\rm ee}=\{I,  \beta_1,  \beta_2,  \beta_1\beta_2\}\times
S, \qquad && \Gamma_{\rm oo}=\gamma^5_1\gamma^5_2\times\Gamma_{\rm
ee},&
\lab{2.5}\\
&\Gamma_{\rm eo}= \{I,  \beta_1,  \im\beta_2,
\im\beta_1\beta_2\}\times\gamma^5_2\times T, \qquad&& \Gamma_{\rm
oe}=\left.\Gamma_{\rm eo}\right|_{1\leftrightarrow2},& \lab{2.6}
\end{alignat}
where $I$ is the unit operator, $S$ and $T$ are sets of scalar and
pseudo-scalar spin factors
\begin{gather}
S = \{S_{(i)},\ i=1,2,3\}\equiv \{I,  \B\sigma_1\cdot\B\sigma_2,
(\B\sigma_1\cdot\B n)(\B\sigma_2\cdot\B n)\}
\lab{2.7}\\
 T = \{T_{(i)},\ i=1,2,3\}\equiv \{\B\sigma_1\cdot\B n,
\B\sigma_2\cdot\B n,  (\B n,\B\sigma_1,\B\sigma_2)\}. \lab{2.8}
\end{gather}
The matrices $\Gamma$ in \re{2.5}, \re{2.6} are grouped as the
even-even operators $\Gamma_{\rm ee}$, odd-odd~$\Gamma_{\rm oo}$,
even-odd $\Gamma_{\rm eo}$ and odd-even operators $\Gamma_{\rm
oe}$, following the Chraplyvy classif\/ication~\cite{Chr53}. All
other $O(3)$-invariant Hermitian matrices built with Dirac
matrices and $\B n$ can be expressed in terms
of~\re{2.5},~\re{2.6}. Correspondingly, the potential~\re{2.3} can
be split into four terms $U_{\rm ee}$, $U_{\rm oo}$, $U_{\rm eo}$,
$U_{\rm oe}$.

    It is convenient to use block-vector representation for the wave function instead of
matrix one as follows
\[
\iPhi(\B r) =\left[\!\!
\begin{array}{cc}
\iPhi_{++}(\B r) & \iPhi_{+-}(\B r) \\
\iPhi_{-+}(\B r) & \iPhi_{--}(\B r) \end{array} \!\!\right] \quad
\dashrightarrow \quad \left[\!\!
\begin{array}{c}
\iPhi_{++}(\B r) \\ \iPhi_{+-}(\B r) \\
\iPhi_{-+}(\B r) \\ \iPhi_{--}(\B r) \end{array} \!\!\right];
\]
here $\iPhi_{++}(\B r)$ is 2$\times$2-spinor matrix which
represents large-large component of wave function \cite{Chr53}
while $\iPhi_{+-}(\B r)$, $\iPhi_{-+}(\B r)$ and $\iPhi_{--}(\B
r)$ are large-small, small-large and small-small components
of~$\iPhi(\B r)$. In this representation the 48 of partial
potentials $U_A(r)$ in \re{2.3} are collected in Dirac multiplets
with common spin factors as follows
\begin{gather}\lab{2.10}
U  =  U_{\rm ee}+U_{\rm oo}+U_{\rm eo}+U_{\rm oe}
  = \sum\limits_{i=1}^{3}
\big(\sD^{(i)} S_{(i)}+ \sO^{(i)} T_{(i)}\big),
\end{gather}
$\sD=\sD_{\rm ee}+\sD_{\rm oo}$, $\sO=\sO_{\rm eo}+\sO_{\rm oe}$
(here we omit a superscript $i$) and
\begin{gather}
\sD_{\rm ee} = \left[\begin{array}{cccc}
U_{11} &&&\\
& U_{22} &&\smash{\lefteqn{\hspace{-1em}\mbox{\Huge 0}}} \\
&& U_{33} &\\
\smash{\lefteqn{\hspace{-0.2em}\mbox{\Huge 0}}}&&& U_{44}
\end{array}\right],
\qquad \sD_{\rm oo} = \left[\begin{array}{cccc}
&&& W_{14}\\
\smash{\lefteqn{\hspace{-0.2em}\mbox{\Huge 0}}}&& W_{23} & \\
& \cc W_{23} &&\\
\cc W_{14} &&&\smash{\lefteqn{\hspace{-1em}\mbox{\Huge 0}}}
\end{array}\right],  \lab{2.11}
\\
\sO_{\rm eo} = \left[\begin{array}{cccc}
0 & W_{12} &&\\
\cc W_{12} & 0 &&\smash{\lefteqn{\hspace{-1em}\mbox{\Huge 0}}} \\
&& 0 & W_{34}\\
\smash{\lefteqn{\hspace{-0.2em}\mbox{\Huge 0}}}&& \cc W_{34} & 0
\end{array}\right],
\qquad \sO_{\rm oe} = \left[\begin{array}{cccc}
&& W_{13} & 0 \\
\smash{\lefteqn{\hspace{-0.2em}\mbox{\Huge 0}}}&& 0 &  W_{24} \\
\cc W_{13} & 0 &&\\
0 & \cc W_{24} &&\smash{\lefteqn{\hspace{-1em}\mbox{\Huge 0}}}
\end{array}\right],
\lab{2.12}
\end{gather}
where matrix entries $U_{\alpha\beta}$ are real while
$W_{\alpha\beta}=X_{\alpha\beta}+\im Y_{\alpha\beta}$ are complex
functions of $r$ (the star ``$\cc{\rule{2ex}{0pt}}$'' denotes a
complex conjugation).


\section{Radial reduction of 2BDE}

    Following  \cite{D-D02,Duv05,Duv06} we choose $\iPhi(\B r)$
to be an eigenfunction of the square $\B j^2$ and the
component~$j_3$ of the total angular momentum $\B j=\B r\times\B
p+\B s = -\im\B
r\times\B\nabla+\tfrac{1}{2}(\B\sigma_1+\B\sigma_2)$ and of the
parity $P$. In a block-matrix representation we have
\begin{gather*}
\iPhi(\B r)= \frac1r\! \left[\!
\begin{array}{c}
\im \phi_1(r)\sY^A(\B n) + \im \phi_2(r)\sY^0(\B n) \\
\phi_3(r)\sY^-(\B n) + \phi_4(r)\sY^+(\B n) \\
\phi_5(r)\sY^-(\B n) + \phi_6(r)\sY^+(\B n) \\
\im \phi_7(r)\sY^A(\B n) + \im \phi_8(r)\sY^0(\B n)
\end{array}
 \!\right] \qquad \mbox{for}\ P=(-)^{j\pm1},
\nn\\
(\sY^A,\sY^0)\leftrightarrow(\sY^-,\sY^+) \qquad  \mbox{for}\
P=(-)^j. \nn
\end{gather*}
Here $\sY^A(\B n)$ is an abbreviation of the 2$\times$2 matrix
bispinor harmonics $\sY^\mu_{\ell sj}(\B n)$ (in conventional
notation \cite{Mes61,Saz86a,KvA04}) corresponding to a singlet
state with a total spin $s=0$ and an orbital momentum $\ell=j$;
harmonics $\sY^0(\B n)$, $\sY^-(\B n)$, $\sY^+(\B n)$ correspond
to triplet with $s=1$ and $\ell=j,j+1,j-1$. Then for $j>0$ the
eigenstate problem \re{2.1} reduces to the set of eight
f\/irst-order dif\/ferential equations with the functions
$\phi_1(r),\dots, \phi_8(r)$ and the energy $E$ to be found. Using
properties of bispinor harmonics (their explicit form is not
important here; see \cite{D-D02,Duv06} for one)
\begin{gather*}
\langle i | k \rangle =\int d\B n \, {\rm Tr}(\sY_{i}^{\dag} \,
\sY_{k})=  \delta_{i\,k},\qquad i,k=A,0,-,+,\\
\B j^2\sY = j(j+1)\sY, \qquad
j =0,1,\dots, \\
j_3\sY = \mu\sY, \qquad
\mu =-j,\dots,j, \\
\B\ell^2\sY = \ell(\ell+1)\sY, \qquad \ell = j, j\pm1,
\\
\B s^2\sY = s(s+1)\sY, \qquad s =
0,1, \\
P\sY^{A,0} = (-)^j\sY^{A,0}, \qquad
P\sY^\mp = (-)^{j\pm1}\sY^\mp, \\
\left[\sY^A\right]^{\rm T} = -\sY^{A}, \qquad
\left[\sY^{0,\mp}\right]^{\rm T} = \sY^{0,\mp},
\end{gather*}
and def\/ining a 8-dimensional vector-function $\sFi(r) =
\{\phi_1(r),\dots, \phi_8(r)\}$ we present this set in the matrix
form
\begin{gather}\lab{3.3}
\left\{\s H(j) \frac{d}{dr} + \s V(r,E,j)\right\}\sFi(r) = 0.
\end{gather}
Here 8$\times$8 matrices $\s H(j)$ and
\begin{gather}\lab{3.4}
\s V(r,E,j)=\s G(j)/r+\s m+\s U(r,j) - E\s I
\end{gather}
possess the properties: $\s H\in$~Re, $\s H^{\rm T}=-\s H$; $\s
V^\dag=\s V$; $\s I$ is 8$\times$8 unity; the diagonal matrix $\s
m$ and matrices $\s H(j)$, $\s G(j)$ are constant (i.e.
independent of $r$), and $\s U(r,j)$ represents the
potential~\re{2.3}. The operator in the l.h.s.\ of~\re{3.3} is
Hermitian with respect to the inner product:
\[
\langle\sPsi|\sFi\rangle=\int_0^\infty dr\,
\big(\sPsi^\dag(r)\sFi(r)\big)
\]
induced by \re{2.4}. In the case $j=0$ components
$\phi_2=\phi_4=\phi_6=\phi_8=0$ so that a dimension of the
problem~\re{3.3} reduces from 8 to 4.

    Uniting bispinor harmonics in two doublets of opposite parity
\[
\of=\left[\begin{array}{c}\sY^A\\\sY^0\end{array}\right],\qquad
\ef=\left[\begin{array}{c}\sY^-\\\sY^+\end{array}\right]
\]
with the properties
\begin{alignat}{3}
& \B\sigma_1\cdot\B\sigma_2\of = \tau\of, \qquad &&
\B\sigma_1\cdot\B\sigma_2\ef = \ef,  &\nonumber \\
& \B\sigma_1\cdot\B n\of = \cf R^{\rm T}\ef, \qquad &&
\B\sigma_1\cdot\B n\ef = \cf R\of,  &\nonumber \\
& \B\sigma_2\cdot\B n\of = -\sigma_3\cf R^{\rm T}\ef, \qquad &&
\B\sigma_2\cdot\B n\ef = -\cf R\sigma_3\of, & \lab{3.7}\\
& (\B\sigma_1\cdot\B n)(\B\sigma_2\cdot\B n)\of = -\sigma_3\ef,
\qquad &&
(\B\sigma_1\cdot\B n)(\B\sigma_2\cdot\B n)\ef = -\cf R\sigma_3\cf R^{\rm T}\of, & \nonumber \\
& (\B n,\B\sigma_1,\B\sigma_2)\of = -2\im\sigma_\uparrow\cf R^{\rm
T}\ef, \qquad && (\B n,\B\sigma_1,\B\sigma_2)\ef = 2\im
R\sigma_\uparrow\ef,&\nonumber
\end{alignat}
where 2$\times$2 matrices $\sigma_\uparrow$, $\sigma_\downarrow$,
$\tau$ and $\cf R$ are def\/ined as follows
\begin{gather*}
\sigma_\uparrow=\tfrac{1}{2}(\cf I+\sigma_3)
=\left[\begin{array}{cc}1&0\\0&0\end{array}\right],\qquad
\sigma_\downarrow=\tfrac{1}{2}(\cf I-\sigma_3)
=\left[\begin{array}{cc}0&0\\0&1\end{array}\right], \nn\\
\tau=\left[\begin{array}{cc}-3&0\\0&1\end{array}\right],\qquad \cf
R=\left[\begin{array}{cc}A&B\\-B&A\end{array}\right]\qquad (\cf
R\in O(2)),\nn
\\
A=\sqrt{\vphantom{\frac{a}{a}}\smash{\frac{j+1}{2j+1}}},\qquad
B=\sqrt{\vphantom{\frac{a}{a}}\smash{\frac{j}{2j+1}}},\qquad
C=\sqrt{j(j+1)},
\end{gather*}
we present all the matrices involved in the equation \re{3.3} in a
block-matrix form
\[
\s m={\rm diag}(m_+\cf I,m_-\cf I,-m_-\cf I,-m_+ \cf I)
\]
(here $m_\pm=m_1\pm m_2$, and $\cf I$ is a 2$\times$2 unity),
\begin{gather*}
\s G=-\left[\begin{array}{cccc} 0 & \cf R(j{+}\sigma_\uparrow) &
\cf R^{\rm T}(j\sigma_3{+}\sigma_\uparrow)
& 0 \\
(j{+}\sigma_\uparrow)\cf R^{\rm T} & 0 & 0 &
(j\sigma_3{+}\sigma_\uparrow)\cf R \\
(j\sigma_3{+}\sigma_\uparrow)\cf R & 0 & 0 &
(j{+}\sigma_\uparrow)\cf R^{\rm
T} \\
0 & \cf R^{\rm T}(j\sigma_3{+}\sigma_\uparrow) & \cf
R(j{+}\sigma_\uparrow)
& 0 \end{array}\right], \nn\\
\s H=\left[\begin{array}{cccc}
0 & -\cf R\sigma_3 & -\cf R^{\rm T} & 0 \\
\sigma_3\cf R^{\rm T} & 0 & 0 & \cf R \\
\cf R & 0 & 0 & \sigma_3\cf R^{\rm T} \\
0 & -\cf R^{\rm T} & -\cf R\sigma_3 & 0
\end{array}\right]\qquad\mbox{for}~P=(-)^{j\pm1},\nn\\
\s G=\left[\begin{array}{cccc} 0 & (j{+}\sigma_\uparrow)\cf R^{\rm
T} & (j\sigma_3{+}\sigma_\uparrow)\cf R
& 0 \\
\cf R(j{+}\sigma_\uparrow)\sigma_3 & 0 & 0 & \cf R^{\rm
T}(j\sigma_3{+}\sigma_\uparrow) \\
\cf R^{\rm T}(j\sigma_3{+}\sigma_\uparrow) & 0 & 0 & \cf
R(j{+}\sigma_\uparrow)\\
0 & (j\sigma_3{+}\sigma_\uparrow)\cf R & (j{+}\sigma_\uparrow)\cf
R^{\rm T}
& 0 \end{array}\right],\\
\s H=\left[\begin{array}{cccc}
0 & -\sigma_3\cf R^{\rm T} & -\cf R & 0 \\
\cf R\sigma_3 & 0 & 0 & \cf R^{\rm T} \\
\cf R^{\rm T} & 0 & 0 & \cf R\sigma_3 \\
0 & -\cf R & -\sigma_3\cf R^{\rm T} & 0
\end{array}\right] \qquad\mbox{for}~P=(-)^{j}.\nn
\end{gather*}

    In order to present the interaction potential
\re{2.10}--\re{2.12} in the block-matrix form we have to calculate
in this representation the form of spin operators \re{2.7},
\re{2.8} using the relations~\re{3.7}. One obtains
\begin{gather*}
\s S ={\rm giag}(\cf S,\cS,\cS,\cf S), \qquad \s T ={\rm giag}(\cf
T,-\cf T^\dag,-\cf T^\dag,\cf T) \qquad\mbox{for}~P=(-)^{j\pm1},
\nn\\
\s S ={\rm giag}(\cS,\cf S,\cf S,\cS), \qquad \s T ={\rm giag}(\cf
T^\dag,-\cf T,-\cf T,\cf T^\dag) \qquad\mbox{for}~P=(-)^{j}, \nn
\end{gather*}
with the following 2$\times$2 matrix blocks $\cf S$, $\cS$, $\cf
T$
$$
\begin{array}{|c|c|c|c|}
\hline
i & 1 & 2 & 3 \\
\hline S_{(i)} & I & \B\sigma_1\cdot\B\sigma_2 &
(\B\sigma_1\cdot\B n)(\B\sigma_2\cdot\B n) \\
\hline
\cf S_{(i)} & \cf I & \tau & -\sigma_3 \\
\hline
\cS_{(i)} & \cf I & \cf I  & -\cf R \sigma_3\cf R^{\rm T}\rule{0pt}{3ex}\\
\hline
\end{array} \qquad
\begin{array}{c}
\begin{array}{|c|c|c|c|}
\hline
i & 1 & 2 & 3 \\
\hline T_{(i)} & \B\sigma_1\cdot\B n & \B\sigma_2\cdot\B n &
(\B n,\B\sigma_1,\B\sigma_2) \\
\hline
\cf T_{(i)} & \cf R & -\cf R\sigma_3 & -2\im\cf R\sigma_\uparrow \\
\hline
\end{array}\\
\rule{0pt}{3ex}
\end{array}
$$

    In the case $j=0$ all 2$\times$2 blocks  in matrices
of the equation \re{3.3} must be replaced by their left upper
entries calculated at $j=0$.


\section{Reduction of 2BDE to a set of second-order ODEs}

    It turns out that ${\rm rank}\,\s H=4$ (2 for $j=0$). In other
words, only four equations of the set~\re{3.3} are dif\/ferential
while remaining ones are algebraic. They can be split by means of
an orthogonal (i.e., from the group $O(8)$ ) transformation
\begin{gather*}
\bar\sFi=\s O\sFi, \qquad \bar{\s H}=\s O\s H\s O^{\rm T} \quad
\mbox{etc}, \qquad  \mbox{where}\quad\s O=\s O_2\s O_1,
\\
\s O_1 =\left[\begin{array}{cccc}
\cf I &&&\\
& \sigma_3\cf R^{\rm T} &&\smash{\lefteqn{\hspace{-1em}\mbox{\Huge 0}}} \\
&& -\cf R^{\rm T}&\\
\smash{\lefteqn{\hspace{-0.2em}\mbox{\Huge 0}}}&&& -\sigma_3
\end{array}\right], \qquad
\s O_2= \frac1{\sqrt{2}} \left[\begin{array}{cccc}
\cf I & 0 & 0 & -\cf I  \\
0 & -\cf I  & \cf I  & 0 \\
0 & \cf I  & \cf I  & 0 \\
\cf I & 0 & 0 & \cf I
\end{array}\right] \qquad \mbox{for} \quad P=(-)^{j\pm1},
\\
\s O_1=\left[\begin{array}{cccc}
\sigma_3\cf R^{\rm T} &&&\\
& \cf I  &&\smash{\lefteqn{\hspace{-1em}\mbox{\Huge 0}}} \\
&& -\sigma_3 &\\
\smash{\lefteqn{\hspace{-0.2em}\mbox{\Huge 0}}}&&& -\cf R^{\rm T}
\end{array}\right],\qquad
\s O_2= \frac1{\sqrt{2}} \left[\begin{array}{cccc}
0 & \cf I  & -\cf I  & 0 \\
\cf I  & 0 & 0 & -\cf I  \\
-\cf I & 0 & 0 & -\cf I  \\
0 & \cf I  & \cf I  & 0
\end{array}\right]
\qquad \mbox{for} \quad P=(-)^{j}.
\end{gather*}
This transformation reduces the matrix $\s H$ to the canonical
form
\begin{gather}\lab{4.4}
\bar{\s H} =2\left[\begin{array}{cc}
\s J & \s 0 \\
\s 0 & \s 0
\end{array}\right],\qquad
\s J =\left[\begin{array}{cc}
0 & \cf I \\
-\cf I &  0
\end{array}\right],
\end{gather}
where $\s J$ is a symplectic (nondegenerate) 4$\times$4 matrix.
Other items of \re{3.3}--\re{3.4} take the form
\begin{gather}\lab{4.5}
\bar{\s G} =2\left[\begin{array}{cccc}
0 & \sigma_\uparrow  & C\sigma_1  & 0 \\
\sigma_\uparrow  & 0 & 0 & 0 \\
C\sigma_1 & 0 & 0 & 0  \\
0 & 0  & 0  & 0
\end{array}\right],\qquad
\bar{\s m} =\left[\begin{array}{cccc}
&&&{\hspace{-0.2ex}m_\pm}\hspace{-0.2ex}\\
\smash{\lefteqn{\hspace{-0.2em}\mbox{\Huge 0}}}
&&{\hspace{-1ex}{-}m_\mp\hspace{-1ex}}& \\
&{\hspace{-1ex}{-}m_\mp\hspace{-1ex}}&&\\
{\hspace{-0.2ex}m_\pm\hspace{-1ex}}&&&\smash{\lefteqn{\hspace{-1em}\mbox{\Huge
0}}}
\end{array}\right]\qquad\mbox{for}\ P=\mp(-)^j,
\\
\lab{4.6} \bar{\s U} =\left[\begin{array}{cccc}
\cf U_{11} & \cf W_{12} & \cf W_{13} & \cf W_{14} \\
\cc{\cf W}_{12} & \cf U_{22} & \cf W_{23} & \cf W_{24} \\
\cc{\cf W}_{13} & \cc{\cf W}_{23} & \cf U_{33} &  \cf W_{34} \\
\cc{\cf W}_{14} & \cc{\cf W}_{24} & \cc{\cf W}_{34} & \cf U_{44}
\end{array}\right];
\end{gather}
here $\cf U_{\alpha\beta}$ are real and~$\cf W_{\alpha\beta}$ are
complex diagonal 2$\times$2 blocks (related linearly to the former
partial potentials $U_{\alpha\beta}$, $W_{\alpha\beta}$ in
\re{2.11}--\re{2.12}) which are dif\/ferent, in general, for the
same $j$ but opposite~$P$. Equivalent amount of real partial
potentials present in \re{4.6} is equal to 32 for $P=(-)^{j\pm1}$
and~32 for $P=(-)^{j}$, but the only 48 among them are
independent.

    Let us present 8-dimensional vectors and
matrices involved in \re{3.3} via 4-dimensional blocks
\[
\bar\sFi =\left[\begin{array}{c} \bar\sFi_1\\ \bar\sFi_2
\end{array}\right], \qquad
\bar{\s V} =\left[\begin{array}{cc}
\bar{\s V}_{11} & \bar{\s V}_{12} \\
\bar{\s V}_{21} &  \bar{\s V}_{22}
\end{array}\right]
\]
and \re{4.4} for $\bar{\s H}$. In these terms the set \re{3.3}
splits into dif\/ferential and algebraic subsets
\begin{gather}
2\s J\bar\sFi'_1+\bar{\s V}_{11}\bar\sFi_1+\bar{\s
V}_{12}\bar\sFi_2=0,
\lab{4.8} \\
\bar{\s V}_{21}\bar\sFi_1+\bar{\s V}_{22}\bar\sFi_2=0. \lab{4.9}
\end{gather}
Eliminating $\bar\sFi_2$ from \re{4.8} by means of \re{4.9} yields
the purely dif\/ferential set for 4-vector $\bar\sFi_1$
\begin{gather}\lab{4.10}
\left\{\s J \frac{d}{dr}+\s V^\bot(r,E,j)\right\}\bar\sFi_1(r)=0.
\end{gather}
The vector $\bar\sFi_2$ is then determined by the algebraic
relation $\bar\sFi_2=-\sL\bar{\s V}_{21}\bar\sFi_1$. Here
\begin{gather}\lab{4.11}
\sL=\left[\bar{\s V}_{22}\right]^{-1}, \qquad \s V^\bot=(\bar{\s
V}_{11}-\bar{\s V}_{12}\sL\bar{\s V}_{21})/2.
\end{gather}

    Next, we present the 4-vector $\bar\sFi_1$ in a 2+2 block form
\begin{gather}\lab{4.12}
\bar\sFi_1(r) = \left[
\begin{array}{c}
\cFi_1 \\ \cFi_2
\end{array}
\right], \qquad \s V^\bot = \left[
\begin{array}{cc}
\cf V_{11} & \cf V_{12} \\
\cf V_{21} & \cf V_{22}
\end{array}
\right], \qquad \sL= \left[
\begin{array}{cc}
\cL_{11} & \cL_{12} \\
\cL_{21} & \cL_{22}
\end{array}
\right],
\end{gather}
then eliminate $\cFi_2$ and arrive at the set of second-order
dif\/ferential equations for 2-vector $\cFi_1$
\begin{gather}\lab{4.13}
{\cf L}(E)\cFi_1 \equiv \left\{\left(\frac{d}{dr}+ \cf
V_{12}\right)\left[\cf V_{22}\right]^{-1} \left(\frac{d}{dr}-\cf
V_{21}\right)+\cf V_{11}\right\}\cFi_1=0.
\end{gather}

    It follows from \re{3.4}, \re{4.5}, \re{4.6}, \re{4.11}, \re{4.12}
that the matrix $\cf V_{22}$ is diagonal. Thus one can perform the
transformation
\[
\cPsi=\cFi_1\left/\sqrt{\cf V_{22}}\right. , \qquad \tilde{\cf L}
(E)=\sqrt{\cf V_{22}}{\cf L}(E)\sqrt{\cf V_{22}},
\]
reducing the equation \re{4.13} to a normal form
\begin{gather}
\tilde{\cf L}(E)\cPsi \equiv  \left\{\left(\frac{d}{dr} + \cf
F\right)\left(\frac{d}{dr} - \cf F^\dag\right) + \cf
Z\right\}\cPsi=0,\lab{4.15}
\end{gather}
where
\begin{gather}
\cf F = \sqrt{\cf V_{22}}\cf V_{12}\left/\sqrt{\cf V_{22}}\right.
\qquad
\mbox{and}\nn\\
\cf Z = \sqrt{\cf V_{22}}\cf V_{11}\sqrt{\cf V_{22}}
+\frac12\left(\cf F\frac{\cf V'_{22}}{\cf V_{22}} + \frac{\cf
V'_{22}}{\cf V_{22}}\cf F^\dag\right) + \frac12\frac{\cf
V''_{22}}{\cf V_{22}} - \frac34\left(\frac{\cf V'_{22}}{\cf
V_{22}}\right)^2.\lab{4.16}
\end{gather}

    If $\cf F$ is Hermitian then the f\/irst-order derivative term
is absent in the operator $\tilde{\cf L}(E)$ and the equation
\re{4.15} takes the matrix two-term form
\begin{gather}\lab{4.18}
\left\{\frac{d^2}{dr^2} + \cf Q\right\}\cPsi=0\qquad \mbox{with}
\cf \quad \cf Q=\cf Z - \cf F^2 - \cf F'
\end{gather}
which is useful for a search of solvable examples of 2BDE.


\section{Solvable oscillator-like models}

    A construction of the matrix  $\cf V_{22}$ \re{4.12}, \re{4.11}
involved in the equations \re{4.10}, \re{4.13}, \re{4.15} and
\re{4.18} is a source of non-physical singularities which very
complicate an analysis of these equations. Moreover, singularities
may make a standard boundary value problem (with constraints at
$r=0,\infty$) incorrect. We note that in the free-particle case
$\cf V_{22}$ is a constant (free of~$r$). Let us require of  $\cf
V_{22}$ to be so in the presence of interaction. Suf\/f\/icient
conditions for this are the equalities
%
\begin{gather}\lab{5.1}
\cf U_{22}=\cf U_{33}=\cf W_{23}=\cf W_{24}=\cf W_{34}=0
\end{gather}
for entries of the matrix \re{4.6} leading to a rather general
potential of the form \re{2.3} with 18~arbitrary real partial
potentials. Four of them compose the even-even part $U_{\rm ee}$
of the interaction, four other constitute the odd-odd $U_{\rm
oo}$, and the remaining 10 ones form potentials of the $U_{\rm
oe}$ and~$U_{\rm eo}$ type. Due to the constraints~\re{5.1} both
the $U_{\rm ee}$ and $U_{\rm oo}$ interaction terms contain the
factor $1-\B\sigma_1\cdot\B\sigma_2$. It cancels the interaction
on a triplet part of the wave function $\iPhi$ and leads to a
dubious bound state spectrum of poor physical meaning. Thus we
discard these potentials by means of the conditions
\[
\cf U_{11}=\cf U_{44}=\cf W_{14}=0
\]
for every of the $P=-(-)^{j}$  and $P=(-)^{j}$ parity cases.

    The following requirement
\begin{gather}\lab{5.3}
\mbox{Re}\,\cf W_{12}=\mbox{Re}\,\cf W_{13}=0
\end{gather}
causes the equality $\mbox{Im}\,\cf F=0$ for \re{4.16} and thus
$\cf F^\dag=\cf F$ (since $\cf F$ is diagonal, due to the
equali\-ties~\re{5.1}). By this we provide the two-term form
\re{4.18} for the wave equation. On the other hand ten partial
potentials contained in the entries $\cf W_{12}$ and $\cf W_{13}$
of the matrix \re{4.6} reduce, due to \re{5.3}, to 6 ones which
form interaction terms of $U_{\rm oe}$ and $U_{\rm eo}$ type.
Among them the term $U_0(r)(\B\alpha_1-\B\alpha_2)\cdot\B n$ with
an arbitrary function $U_0(r)$ is purely gauge and can be
compensated by the phase factor $\exp\left\{-\im\int
drU_0(r)\right\}$ of $\sPsi$. The resulting anzatz for a potential
is as follows
\begin{gather*}
U
=\im\left\{U_1(r)+U_2(r)\gamma^5_1\gamma^5_2\right\}\beta_1\beta_2
(\B\alpha_1{-}\B\alpha_2)\cdot\B n +
U_3(r)\left(\gamma^5_1+\gamma^5_2\right)
(\B\sigma_1\times\B\sigma_2)\cdot\B n \nn\\
\phantom{U=}{}+\im
\left\{U_4(r)\beta_1\gamma^5_1-U_5(r)\beta_2\gamma^5_2\right\}
(\B\sigma_1-\B\sigma_2+\im\B\alpha_1\times\B\alpha_2) \cdot\B n,
\nn
\end{gather*}
where $U_1(r),\dots,U_5(r)$ are arbitrary real functions.

    After aforementioned simplif\/ications the matrices $\cf Q_+$ and
$\cf Q_-$ involved in the equation~\re{4.18} and referred to the
$P=-(-)^{j}$ and $P=(-)^{j}$ parity cases respectively have the
form
\begin{gather*}
\cf Q_\pm=\frac14\left(E-\frac{m_+^2}E\right)\left(E-\frac{m_-^2}E\right)\\
\phantom{\cf Q_\pm=}{} -
\left[f_{\uparrow\pm}^2+\left(\frac{d}{dr}+\frac2r\right)f_{\uparrow\pm}
-
\frac{m_\mp}E\left(2f_{\uparrow\pm}+\frac{d}{dr}+\frac2r\right)u_\pm+u_\pm^2
\right]\sigma_\uparrow
\\
\phantom{\cf Q_\pm=}{}
-\left[f_{\downarrow}^2\pm\frac{d}{dr}f_{\downarrow}\right]\sigma_\downarrow
- \frac Cr\left[\frac{m_\mp}E\left(f_{\uparrow\pm}\pm
f_{\downarrow}\right)-u_\pm\right]\sigma_1 -\frac{C^2}{r^2},
\end{gather*}
where f\/ive functions
\[
f_{\uparrow\pm}=\mp(U_1+U_2)-2U_3,\qquad
f_{\downarrow}=U_2-U_1,\qquad u_\pm=-2(U_4\pm U_5)
\]
represent equivalently partial potentials. They are a convenient
choice when looking for solvable models.

    {\bf Models I.} The choice $u=0$ and $f^2=a^2r^2$, where $a={\rm const}$,
leads to four possibilities
\begin{alignat*}{3}
& {} \mbox{Ia}) \quad
f_{\uparrow+}=-f_{\uparrow-}=-f_\downarrow=-ar, \qquad &&
\mbox{Ic}) \quad f_{\uparrow+}=f_{\uparrow-}=f_\downarrow=-ar,& \\
& {} \mbox{Ib}) \quad
f_{\uparrow+}=-f_{\uparrow-}=f_\downarrow=-ar, \qquad &&
\mbox{Id}) \quad f_{\uparrow+}=f_{\uparrow-}=-f_\downarrow=-ar, &
\end{alignat*}
which correspond to the following sets of partial potentials (only
non-zero ones are shown)
\begin{alignat*}{3}
& \mbox{Ia}) \quad U_2=ar, \qquad &&
\mbox{Ic}) \quad U_1=-U_2=U_3=\tfrac{1}{2} ar, & \\
& \mbox{Ib}) \quad U_1=ar, \qquad && \mbox{Id}) \quad
U_1=-U_2=-U_3=-\tfrac{1}{2} ar.&
\end{alignat*}
All cases lead to the oscillator-like equation
%
\begin{gather}\lab{5.9}
\left\{\frac{d^2}{dr^2} +
\frac14\left(E-\frac{m_+^2}E\right)\left(E-\frac{m_-^2}E\right) -
a^2r^2 - \frac{C^2}{r^2} + a\cf D_\pm\right\}\cPsi=0,
\end{gather}
where $\cf D_\pm$ is a constant 2$\times$2 matrix. In the case Ia)
this matrix is diagonal,
\begin{gather}
\mbox{Ia}) \quad \cf
D_\pm=\,\mbox{diag}\,\{\delta_{\uparrow\pm},\delta_{\downarrow\pm}\},
\lab{5.10}\\
\phantom{\mbox{Ia}) \quad}{} \delta_{\uparrow\pm}=\pm3,\quad
\delta_{\downarrow\pm}=\mp1  \qquad \mbox{for} \ P=\mp(-)^j.
\lab{5.11}
\end{gather}
The energy spectrum can be found from the algebraic relation
\begin{gather}\lab{5.12}
\frac14\left(E-\frac{m_+^2}E\right)\left(E-\frac{m_-^2}E\right) =
|a|(2n+3) - a\delta_{\updownarrow\pm} , \qquad n=j+2n_r,
\end{gather}
where $n_r=0,1,\dots$ is a radial quantum number. Given $j$ and
$n_r$ we have 4 equations determining 4 positive values of energy
$E_{\updownarrow\pm}$
\begin{gather}\lab{5.13}
E_{\updownarrow\pm}=\sum_{a=1}^{2}\sqrt{m_a^2+
|a|(2n+3)-a\delta_{\updownarrow\pm}}.
\end{gather}

    In case Ib) the matrix $\cf D_\pm$ is not diagonal,
\begin{gather}\lab{5.14}
\cf D_\pm = \pm\left[\begin{array}{cc}
3 & -2m_\mp C/E \\
-2m_\mp C/E & 1
\end{array}\right] \qquad \mbox{for} \ P=\mp(-)^j.
\end{gather}
It can be reduced, by means of $O(2)$-transform, to the form
\re{5.10} with eigenvalues
\begin{gather}\lab{5.15}
\mbox{Ib})\quad \delta_{\uparrow\pm} = 2 + \sqrt{1+(m_\mp
C/E)^2},\!\qquad \delta_{\downarrow\pm} = 2 - \sqrt{1+(m_\mp
C/E)^2} \!\qquad \mbox{for}\ P=\mp(-)^j.\!\!\!\!
\end{gather}
The set \re{5.9} splits into two oscillator-like equations. Since
$\delta$'s \re{5.15} are energy-dependent, the spectral conditions
\re{5.12} appear in this case as irrational equations which can be
reduced to a~fourth-order algebraic equations (third-order if
$m_1=m_2$) with respect to $E$.

    In cases Ic) and Id) the matrix $\cf D_\pm$
is equal to either \re{5.10}--\re{5.11} or \re{5.14} for $P=(-)^j$
case and to another (complementary) one for opposite parity. The
corresponding eigenvalues
\begin{gather*}
\mbox{Ic}) \quad \delta_{\updownarrow+} = 2 \pm
\sqrt{1+(m_-C/E)^2},\qquad \delta_{\updownarrow-}=1\pm2,
\nn\\
\mbox{Id}) \quad \delta_{\updownarrow+}=1\pm2,\qquad
\delta_{\updownarrow-} = 2 \pm \sqrt{1+(m_+C/E)^2} \nn
\end{gather*}
are to be used in the spectral formulae \re{5.12} or \re{5.13}.

\begin{figure}[t]
\centerline{\includegraphics[width=15.5cm]{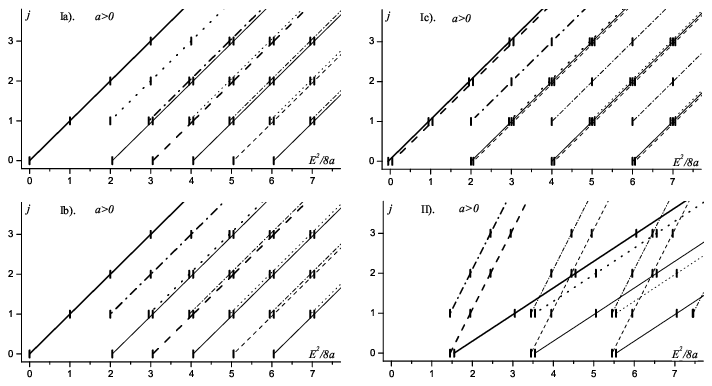}}

\rightline{\includegraphics[width=5cm]{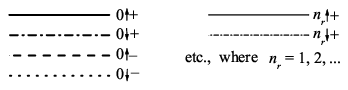}}


\vspace{-1.5cm}

\begin{tabular}{p{7.0cm}}
\caption{Regge trajectories from massless models I and
II.}\label{fig1}
\end{tabular}
\vspace{-0.5cm}
\end{figure}

        If $m_1=m_2=0$, the constants
$\delta_{\updownarrow\pm}$ do not depend on $E$ and $j$ and the
equation \re{5.12} simplif\/ies to the explicit formula for the
energy spectrum
\[
E^2_{\updownarrow\pm}=4[|a|(2n+3)-a\delta_{\updownarrow\pm}].
\]
In this case energy levels $E_{\updownarrow\pm}$ build up in the
($E^2,j$)--plane into parallel straight lines (so called {\em
Regge trajectories}) with the slope rate $8|a|$. All the
trajectory can be labeled unambiguously by the triplet of quantum
numbers ${n_r}$, ${\updownarrow}$, ${\pm}$ but some trajectories
corresponding to dif\/ferent numbers may coincide (i.e.,
degenerate). Examples of Regge trajectories for massless models
Ia)--Id) are shown in the Fig.~\ref{fig1}.

    {\bf Models II.} The choice $f_\uparrow=0$ and $f_\downarrow^2=u^2=a^2r^2$
leads to four possibilities
\begin{alignat*}{3}
& \mbox{IIa}) \quad f_\downarrow=\pm u_\pm=-ar, \qquad &&
\mbox{IIc}) \quad f_\downarrow=u_\pm=-ar, & \\
& \mbox{IIb}) \quad f_\downarrow=\mp u_\pm=-ar, \qquad &&
\mbox{IId}) \quad f_\downarrow=-u_\pm=-ar,&
\end{alignat*}
which correspond to the following sets of partial potentials (only
non-zero ones are shown)
\begin{alignat*}{3}
& \mbox{IIa}) \quad U_1=U_4=\tfrac{1}{2} ar, \qquad &&
\mbox{IIc}) \quad U_1=U_5=\tfrac{1}{2} ar, & \\
& \mbox{IIb}) \quad U_1=-U_4=-\tfrac{1}{2} ar, \qquad &&
\mbox{IId}) \quad U_1=-U_5=-\tfrac{1}{2} ar.&
\end{alignat*}
This choice leads to four solvable models described by the
oscillator-like equation~\re{5.9} with some matrices $\cf D$. Here
we consider the massless case only. All models IIa)--IId) have the
same spectrum
\begin{alignat*}{3}
& a>0\, : \qquad && a<0\, : & \\
& E^2_{\uparrow+}=4a(3j+4n_r+3),  \qquad && E^2_{\uparrow+}=4|a|(j+4n_r+3), & \\
& E^2_{\downarrow+}=4a(j+4n_r+2), \qquad && E^2_{\downarrow+}=4|a|(3j+4n_r+4), & \\
& E^2_{\uparrow-}=4a(j+4n_r+3), \qquad && E^2_{\uparrow-}=4|a|(3j+4n_r+3), & \\
& E^2_{\downarrow-}=4a(3j+4n_r+4), \qquad &&
E^2_{\downarrow-}=4|a|(j+4n_r+2).&
\end{alignat*}
In contrast to models I, here the Regge trajectories are not
parallel: two families are sloped by~$4|a|$, two other -- by
$12|a|$ (see Fig.~\ref{fig1}).

    {\bf Models III.} The conditions $f_\uparrow-=u_+=0$,
$f_{\uparrow+}^2=f_{\downarrow}^2=u_-^2=a^2r^2$ yield four
solvable models one example of which corresponds to the potentials
\[
U_1=\mbox{\small$\frac{3}{4}$}ar,\qquad U_2=U_5=-\qu ar, \qquad
U_3=U_4=\qu ar.
\]
These models (in the massless case) lead to sequences of Regge
trajectories with three dif\/ferent slopes: $4|a|$,  $8|a|$
and~$12|a|$.

  Two of the above models are known in the literature.
Model Ia) is equivalent to the Sazdjian pseudo-scalar
conf\/inement model~\cite{Saz86a}; it coincides with one of
versions of two-body Dirac oscillator proposed in~\cite{M-M94}.
Supersymmetric aspect of these models has been studied
in~\cite{Saz88,MQS95}. Model Ib) is a generalization of another
version of Dirac oscillator \cite{MLV91} to the case of
dif\/ferent rest masses.

        While having no direct physical meaning, the energy
eigenvalues of the models possess some important features of
actual meson spectra which we discuss in the next section.


\section{Light meson spectra}

    Characteristic features of mass spectra of light mesons (consisting
of u, d and s quarks) can be summarized roughly in the following
idealized picture \cite{B-P91,GCO90,Duv05,Duv06}:
\begin{enumerate}\itemsep=0pt
\item Meson states are clustered in the family of straight lines
in the ($E^2,j$)--plane known as Regge trajectories. \item
    Regge trajectories are parallel; slope parameter $\omega$
is an universal quantity, $\omega=1.15\;{\rm GeV}^2$. \item
    Mesons can be classif\/ied non-relativistically, as $\left(n^{2s+1} \ell_j\right)$-states
of quark-antiquark system ($n=n_r+\ell+1$ where $\ell$ and $n_r$
are the orbital and radial quantum number). \item
    Spectrum is $\ell$$s$-degenerated, i.e., masses are distinguished by
$\ell$ (not by $j$ or $s$) and $n_r$. \item
    States of dif\/ferent $\ell$ and $n_r$ possess an accidental
degeneracy which causes a tower structure of the spectrum.
\end{enumerate}

\begin{figure}[t]

\centerline{\includegraphics[width=15.5cm]{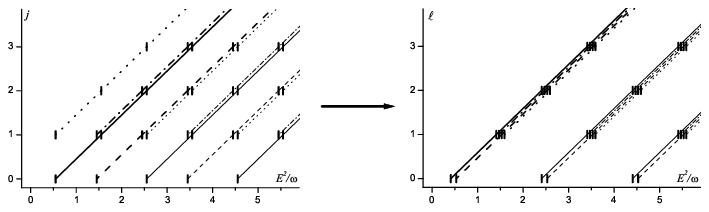}}

\rightline{\includegraphics[width=5cm]{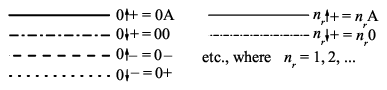}}
\vspace{-1.4cm}

\begin{tabular}{p{7cm}}
\caption{Regge trajectories of idealized light meson spectra;
$\kappa=2$, $\zeta=1/2$.}\label{fig2}
\end{tabular}
\vspace{-0.5cm}
\end{figure}

It follows from items 1--3 that there exist 4 principal ($n_r=0$)
trajectories, one of which includes singlet states ($s=0$,
$j=\ell$), and three others collect dif\/ferent triplet states
($s=1$, $j=\ell,\ell\pm1$). Each of principal trajectories heads a
sequence of daughter trajectories ($n_r=1,2,\dots$). Item~4 means
that in the ($E^2,\ell$)-plane the four parents degenerate
completely (this is concerned also with daughters of the same
$n_r$). Then energy levels of q-\=q system can be described by
a~formula
\begin{gather}\lab{6.1}
E^2\approx\omega(\ell + \kappa n_r + \zeta),
\end{gather}
where a constant $\zeta$ depends on a f\/lavor content of mesons
($\zeta\approx1/2$ for ($\pi$-$\rho$)-family of mesons; it  grows
together with quark masses). Finally, item~5 restricts the
constant $\kappa$ to an integer or rational number \cite{GCO90}.
One frequently puts $\kappa=2$ (oscillator-like degeneracy)
\cite{K-N73,Tak79,I-O94,Sim94}, rarely $\kappa=1$ (Coulomb-like
one) \cite{Khr87,Duv01}. Fig.~\ref{fig2} illustrates the case of
($\ell{+}2n_r$)-degeneracy where corresponding Regge trajectories
are plotted in planes ($E^2,j$) and ($E^2,\ell$).

    Actual spectra of light mesons dif\/fer from the idealized
spectrum~\re{6.1}. A f\/inite number of mesons are known, some
meson states are ambiguously identif\/ied by quantum numbers,
Regge trajectories are not quite straight (especially, in their
bottom) \cite{I-S01} degeneracy is approximated ($\sim5\div6\%$ of
$\omega$) etc. Thus the equation~\re{6.1} represents light meson
spectra as a rather rough approximation, and its subsequent
correction is implied.

    Most of features described above are characteristic of spectra
of the models Ia)--Id) (see Section~5). Indeed, in the massless
case Regge trajectories are straight and parallel.  The
trajectories $\uparrow$$+$ consisting of energy levels
$E_{\uparrow+}$ of parity $P=-(-)^j$ can be treated as a singlet
trajectories corresponding to $s=0$, $\ell=j$ (and labeled by
``A''; see Section~3). Similarly, triplet quantum numbers $s=1$,
$\ell=j,j\mp1$ can be prescribed to the trajectories
$\downarrow$$+$, $\updownarrow$$-$ (labeled by~``0'', ``$\mp$'').
In these terms the spectrum of the equal-mass ($m_1=m_2\equiv m$)
model Ia) is as follows:
\[
E^2=8|a|(\ell+2n_r+3/2)+8a(2s-3/2) + 4m^2
\]
which agrees with results presented in
\cite{Saz86a,Saz88,M-M94,MQS95}. For the model Ic) we have
\[
E^2=8|a|(\ell+2n_r+3/2)+8a(s-3/2) + 4m^2
\]
(the only dif\/ference consists in a coef\/f\/icient at $s$).

    All models Ia)--Id) reveal degenerated trajectories. But
the degeneracy is not a $\ell s$ one. In order to provide $\ell
s$-degeneracy one should have a model possessing additional
$O(3)$-symmetry with corresponding conserved
generators~\cite{BPP88}. The total spin $\B
s=\tfrac{1}{2}(\B\sigma_1+\B\sigma_2)$ does not match for this
role since even the free two-body Dirac Hamiltonian does not
commute with $\B s$. Below we propose generalizations of the
I-type models which reveals an approximated $\ell s$-degeneracy at
some values of free parameters.

    {\bf Model IV} is an integrable
extension of the model Ib). We put $u=0$ and
\begin{gather}\lab{6.4}
f_{\uparrow\pm}=\mp ar+(\varkappa_\pm-1)/r, \qquad
f_\downarrow=-ar+\chi/r,
\end{gather}
where $\varkappa_+$, $\varkappa_-$ and $\chi$ are arbitrary
constants. So the model includes, together with a slope parameter
$a$ and rest particle masses $m_1$, $m_2$, six arbitrary
parameters.

 The choice \re{6.4} corresponds to the following partial
potentials
\begin{gather}\lab{6.5}
U_1=ar-\frac{\varkappa_+-\varkappa_-+2\chi}{4r}, \qquad
U_2=-\frac{\varkappa_+-\varkappa_--2\chi}{4r}, \qquad
U_3=-\frac{\varkappa_++\varkappa_--2}{4r}.
\end{gather}
The matrix $\cf Q_\pm$ in the wave equation \re{4.18} has the form
\begin{gather}
\cf
Q_\pm=\frac14\left(E-\frac{m_+^2}E\right)\left(E-\frac{m_-^2}E\right)-
a^2r^2+a(2\chi\pm1)-\frac{C^2+\chi(\chi\pm1)}{r^2}
\nn\\
\phantom{\cf Q_\pm=}{}+2\left(\pm
a-\frac{\varkappa_\pm\pm\chi-1}{2r^2}\right) \left[\displaystyle{
\begin{array}{cc}
\varkappa_\pm\mp\chi & -m_\mp C/E \\
-m_\mp C/E & 0\rule{0ex}{4ex}
\end{array}}\right].\lab{6.6}
\end{gather}
The matrix in r.h.s.\ of \re{6.6} and thus the $\cf Q_\pm$ itself
can be diagonalized, similarly to case~Ib). Then the equation
\re{4.18} splits into two oscillator-like equations with $j$- and
$E$-dependent free term and coef\/f\/icient at $1/r^2$ term.
Spectral conditions which in general are cumbersome irrational
equations cannot be solved explicitly. In a massless case we get
exact expressions for Regge trajectories
\begin{gather*}
E^2_{\uparrow\pm}=8|a|(k_{\uparrow\pm}+2n_r+\se) \mp
8a(\varkappa_\pm+\tfrac{1}{2}), \qquad
k_{\uparrow\pm}=\sqrt{(j+\tfrac{1}{2})^2+\varkappa_\pm(\varkappa_\pm-1)}-\tfrac{1}{2}, \\
E^2_{\downarrow\pm}=8|a|(k_{\downarrow\pm}+2n_r+\se) -
8a(\chi\pm+\tfrac{1}{2}),\qquad
k_{\downarrow\pm}=\sqrt{(j+\tfrac{1}{2})^2+\chi(\chi\pm1)}-\tfrac{1}{2}.
\end{gather*}

    The Coulomb-like terms in the potentials \re{6.5} bend Regge
trajectories (especially in the bottom) leaving their asymptotics
(at $j\to\infty$) rectilinear (since
$k_{\updownarrow\pm}\stackrel{j\to\infty}{\approx} j + O(1/j)$).
When choosing $\varkappa_+=\chi$, $\varkappa_-=-\chi-2$ (where
$\chi$ is still arbitrary), trajectories become degenerated
asymptotically in the plane ($E^2,\ell$). This model satisf\/ies
approximately item~4 of the properties of meson spectra but leads
to top-heavy meson masses.

    {\bf Model V} is another extension of Ib) with the same
amount of free parameters as in model~IV. We put
\[
u_\pm=2\varkappa_\pm/r,\qquad
f_{\uparrow\pm}=\mp\{ar-(\chi\mp1)/r\}, \qquad
f_\downarrow=-ar+\chi/r,
\]
which correspond to the potentials
\[
U_1=ar-\frac{\chi}{r}, \qquad U_2=0, \qquad U_3=\frac{1/2}{r},
\qquad U_4=-\frac{\varkappa_++\varkappa_-}{2r}, \qquad
U_5=-\frac{\varkappa_+-\varkappa_-}{2r}.
\]
The matrix $\cf Q_\pm$ is as follows
\begin{gather*}
\cf
Q_\pm=\frac14\left(E-\frac{m_+^2}E\right)\left(E-\frac{m_-^2}E\right)-
a^2r^2+a(2\chi\pm1)-\frac{C^2+\chi(\chi\pm1)}{r^2}
\nn\\
\phantom{\cf
Q_\pm=}{}-2\left\{\pm\frac{m_\mp}{E}\left(a-\frac{\chi\mp1/2}{r^2}\right)+
\frac{\varkappa_\pm}{r^2}\right\}\left[\displaystyle{
\begin{array}{cc}
2\varkappa_\pm & C \\
C & 0\rule{0ex}{4ex}
\end{array}}\right].
\nn
\end{gather*}
Again, an evident diagonalization of $\cf Q_\pm$ leads to a split
pair of oscillator-like equations and spectral conditions similar
to case IV. In a massless case we have the spectrum
\begin{gather*}
E_{\updownarrow}^2=8|a|(k_{\updownarrow}+2n_r+\se)-8a(\chi\pm\tfrac{1}{2}),
\end{gather*}
where
\begin{gather*}
k_{\updownarrow}=\sqrt{C^2+(\chi\mp\tfrac{1}{2})^2
+2|\varkappa|(|\varkappa|\pm\sqrt{C^2+\varkappa^2})}-1/2
\stackrel{j\to\infty}{\approx} j\pm|\varkappa| + O(1/j)
\end{gather*}
and parity indexes ``$\pm$'' are implied. Regge trajectories are
asymptotically linear but there is no a choice of parameters
providing the $\ell$$s$-degenegacy.

Models IV and V represent maximal solvable extensions of model Ib)
found in this work. There exist extensions of other model of
family~I which ref\/lect several features of light meson spectra
but are exactly solvable with less number of free parameters than
models IV and V. Here we consider one example which has close
relevance to a meson spectroscopy.

    {\bf Model VI} is the integrable extension of model Ic).
We put $u=0$ and
\[
f_{\uparrow+}=-ar+(\varkappa-1)/r, \qquad
f_{\uparrow-}=f_\downarrow=-ar+\chi/r,
\]
which corresponds to the potentials
\begin{gather*}
U_1=\frac12\left(ar-\frac{\varkappa+\chi-1}{2r}\right), \qquad
U_2=-\frac12\left(ar+\frac{\varkappa-3\chi-1}{2r}\right), \\
U_3=\frac12\left(ar+\frac{\varkappa+\chi-1}{2r}\right).
\end{gather*}
The matrix $\cf Q_+$ is identical to that \re{6.6} of model IV and
a treatment is the same. The matrix~$\cf Q_-$ in the present case
is diagonal
\begin{gather*}
\cf Q_-=\mbox{diag}\,\{Q_{\uparrow-},Q_{\downarrow-}\},
\\
Q_{\updownarrow-}=\frac14\left(E-\frac{m_+^2}E\right)\left(E-\frac{m_-^2}E\right)-
a^2r^2+a\{2(\chi\pm1)+1\}-\frac{C^2+\chi(\chi+1)}{r^2}.
\end{gather*}

\begin{figure}[t]
\centerline{\includegraphics[width=8cm]{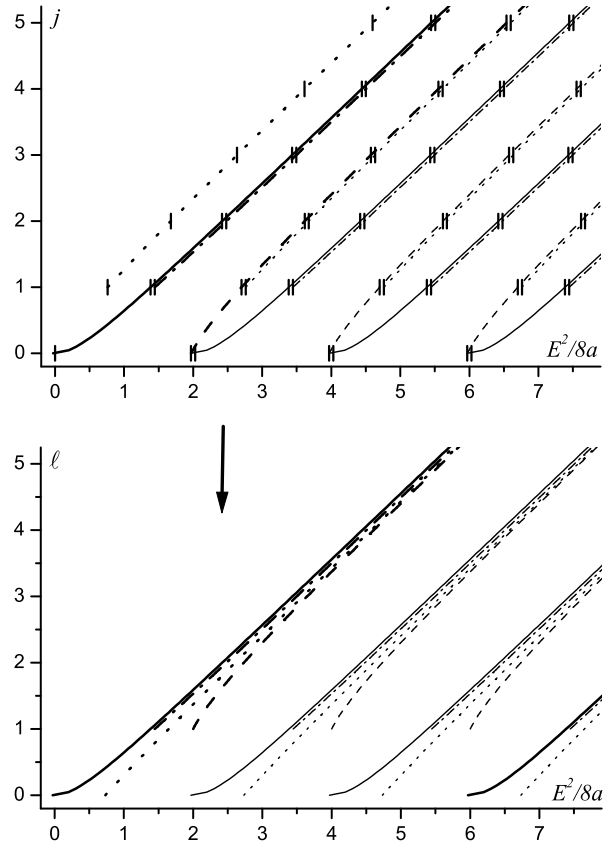}}
\caption{Regge trajectories from model VI; $m_1=m_2=0$,
$\chi=\varkappa=1/2$.} \label{fig3}\vspace{-0.2cm}
\end{figure}

    In the case of equal particle masses, $m_1=m_2\equiv m$, we get the energy
spectrum explicitly{\samepage
\begin{gather}
E^2_{\uparrow+}=8|a|(k_{\uparrow+}+2n_r+\se) -
8a(\varkappa+\tfrac{1}{2}) + 4m^2, \qquad
k_{\uparrow+}=\sqrt{(j+\tfrac{1}{2})^2+\varkappa(\varkappa-1)}-\tfrac{1}{2},
\nn\\
 E^2_{\downarrow+}=8|a|(k_{\downarrow+}+2n_r+\se) - 8a(\chi+\tfrac{1}{2}) +
4m^2, \qquad
k_{\downarrow+}=\sqrt{(j+\tfrac{1}{2})^2+\chi(\chi-1)}-\tfrac{1}{2},\!\!\!\lab{6.16}
\\
 E^2_{\updownarrow-}=8|a|(k_{\updownarrow-}+2n_r+\se) -
8a(\chi+\tfrac{1}{2}\pm1) + 4m^2, \qquad
k_{\updownarrow-}=\sqrt{(j+\tfrac{1}{2})^2+\chi(\chi+1)}-\tfrac{1}{2}.\nn
\end{gather}
Regge trajectories are asymptotically linear and, if
$\varkappa=\chi$, $\ell$$s$-degenerated (see Fig.~\ref{fig3}).}

    Let us try to describe the spectrum of light mesons of the
($\pi$-$\rho$)-family by means of equations~\re{6.16} using four
arbitrary parameters $a$, $\chi$, $\varkappa$ and $m$ as
adjustable parameters. We note that the intersection of the
$\rho(770)$-trajectory (which is 0$\downarrow$$-$ in our terms)
with the $E^2$-axis in the ($E^2,j$)-plane is negative while it
follows from~\re{6.16}: $E^2_{\downarrow-}(j{=}0,n_r{=}0)\ge4m^2$.
On the whole one can achieve a qualitative and partially numerical
agreement of the model with experimental data when supposing that
the meson masses squared $M^2$ are related to $E^2$ as follows:
$M^2=E^2-c^2$ where the parameter $c^2>0$ is common for all
states\footnote{The parameter $c^2$ absorbs in \re{6.16} the term
$4m^2$ which then is set to zero.}. Similar picture arises within
other models, in particular, in \cite{S-C93} where the meaning of
the constant $c$ is discussed. Alternatively, one supposes $m^2<0$
(instead of the use of $c$) in some SROM potential models
\cite{Tak79,I-Y87}. This is not acceptable here since imaginary
rest masses break the Hermicity of the two-body Dirac
Hamiltonian~\re{2.1}.

\begin{figure}[t]
\centerline{\includegraphics[width=8cm]{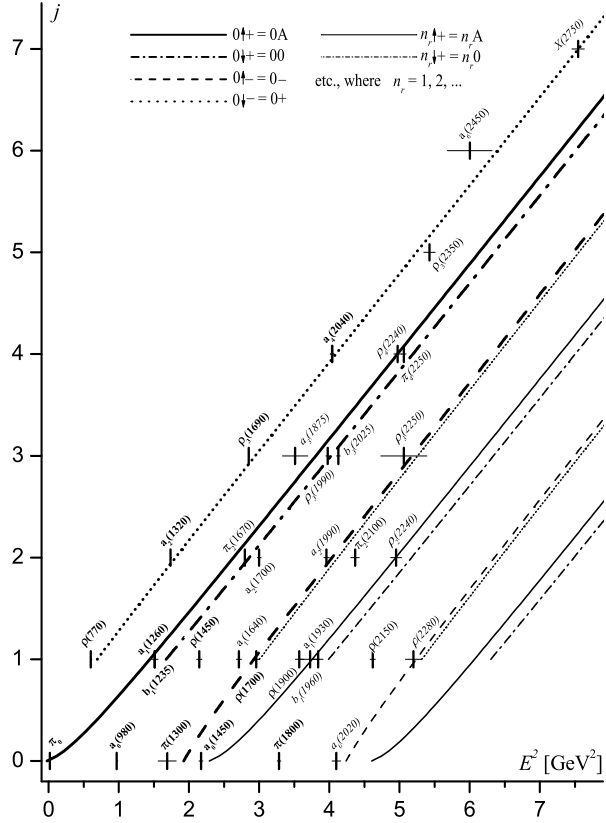}}

\caption{Spectrum of $\pi$- and $\rho$-mesons and optimal Regge
trajectories from model VI. Reliable data are inscribed in {\bf
bold}, doubtful -- in {\em italic}. Thin horizontal lines denote
measurement errors.}\label{fig4} \vspace{-0.2cm}
\end{figure}

    It is shown in Fig.~\ref{fig4} the spectrum of the ($\pi$-$\rho$)-family
plotted with the data \cite{PDG06} and corresponding Regge
trajectories built with optimal values of parameters:
$a=0.145~\mbox{GeV}^2$, $\chi=0.227$, $\varkappa=0.342$,
$c^2=0.346~\mbox{GeV}^2$. Principal trajectories f\/it majority of
meson states. Some radially excited levels, however, are situated
far from daughter trajectories.

    In the case $\varkappa=\chi=0$  trajectories become linear and
degenerated exactly, and formulae \re{6.16} drastically simplify
\[
E^2=8|a|(\ell+2n_r+\se) + 4(m^2-a).
\]
Except for a def\/inition of an additive constant this formula
describes exactly the spectrum of the relativistic system of two
spinless particles harmonically bound (SROM)
\cite{K-N73,Tak79,I-O94}. In the present case this simplicity is
achieved owing to a rather nontrivial choice of the total
potential
\begin{gather*}
U =\frac{\im}2ar\beta_1\beta_2(\gamma^5_1-\gamma^5_2)
(\B\sigma_1+\B\sigma_2)\cdot\B n
+\frac{\im}{4r}\beta_1\beta_2(\gamma^5_1+\gamma^5_2)
(\B\sigma_1-\B\sigma_2)\cdot\B n \nn\\
\phantom{U=}{}+\frac12\left(ar+\frac1{2r}\right)(\gamma^5_1+\gamma^5_2)
(\B\sigma_1\times\B\sigma_2)\cdot\B n. \nn
\end{gather*}

    An exact $\ell$$s$-degeneracy indicates the existence
of additional $O(3)$-symmetry which cannot be generated by
components of the total spin $\B
s=\tfrac{1}{2}(\B\sigma_1+\B\sigma_2)$. A search of relevant
conserved quantities is beyond the scope of this work.


\section{Summary}

    In this work we consider a two-body Dirac equation with
general local (in the position representation) potential found by
Nikitin and Fushchich in \cite{N-F91}. It is parameterized by 48
real functions of $r$ and is presented here in a matrix form.

    Owing to $O(3)$-invariance of 2BDE it is reduced (via a
separation of variables) to a set of eight f\/irst-order ODEs for
radial components of the wave function. Then using a chain of
transformations we eliminate some components of the wave function
in favor of other ones and arrive at the related pair of
second-order ODEs. Coef\/f\/icients of these equations have, in
general, poles at some energy-dependent points $r_E$ which are
absent in an original potential. These singularities may
complicate an analysis and a solution of the problem.

    The structure of second-order reduction of general 2BDE suggests a wide
class of potentials (parameterized by 14 arbitrary functions) for
which the problem is free of non-physical singulari\-ties. 
Within this class a family of exactly solvable models
is found which generalize known two-particle
Dirac oscillators \cite{Saz86a,Saz88,MLV91,M-M94,MQS95}.
In particular,
two of these (models~VI and~V) are 6-parametric integrable
extensions of the oscillator model~\cite{MLV91}. Regge
trajectories following from these models have parallel rectilinear
asymptotes but are curved in their lover segments.

    The 5-parametric model VI is used as a solvable potential
model of light mesons. Special choice of parameters leads to
linear Regge trajectories which possess an exact
$\ell$$s$-degeneracy and the accidental (${j{+}2n_r}$)-degeneracy.
In other words, the model restores the idealized meson spectra
generating by SROM \cite{K-N73,Tak79,I-O94}. The corresponding
two-fermion interaction potential turns out to be surprisingly
intricate.

    Unfortunately, the model fails to describe properly the spectrum
of lightest mesons as it overestimates meson masses squared $M^2$
by certain (common for all states) constant $c^2$. Up to this
discrepancy the description of ($\pi$-$\rho$)-family is adequate
and has been realized. The choice of parameters in this case
dif\/fers slightly from that of the completely degenerated model
so that the Regge trajectories are somewhat curved in the bottom.
The f\/it is good for most of orbital meson excitations and worse
-- for radial ones due to shortage of daughters following from the
model. It follows from this fact that the accidental
(${j{+}2n_r}$)-degeneracy inherent approximately or exactly in
this and many other models \cite{K-N73,Tak79,I-O94,Sim94} is less
adequate to actual meson data than the degeneracy of
($j{+}n_r$)-type.

    Despite  the 2BDE used in this work being $O(3)$-invariant and not
truly relativistic equation, it can be considered as some
covariant equation reduced in the center-of-mass frame of
reference. In Appendix we constructed explicitly the
Poincar\'e-invariant equation for 2BDE with potential of general
form and proved its unambiguity.


\pdfbookmark[1]{Appendix. Covariant form of 2BDE}{appendix}
\section*{Appendix. Covariant form of 2BDE}
\renewcommand{\theequation}{A.\arabic{equation}}
\setcounter{equation}{0}

    The two body Dirac equation considered in this work is $O(3)$-invariant
but not Poincar\'e-invariant. Below we construct the manifestly
covariant equations reduction of which in the center-of-mass
reference frame restores the 2BDE~\re{2.1} with arbitrary
potential of the form~\re{2.3}.

    We start with a free-particle system. Following by the
constraint formalism \cite{Saz86,M-S94,M-S95,KvA87,KvA90,KvA99} it
is described by the pair of covariant Dirac equations
\begin{gather}\lab{A.1}
(\gamma_a\cdot p_a - m_a)\iPhi=0, \qquad a=1,2,
\end{gather}
where $\gamma_a\cdot p_a\equiv\gamma_a^\mu p_{a\mu}$, the particle
4-momenta $p_{a\mu}$ ($a=1,2$; $\mu=0,\dots,3$) are conjugated to
particle positions $x_a^\mu$, and the timelike Lorentz metrics
$\|\eta_{\mu\nu}\|={\rm diag}(+,-,-,-)$ is used.

    Before introducing an interaction it is convenient to represent
the equations \re{A.1} in a collective form. For this purpose we
perform the canonical transformation $\left(x_a^\mu,
p_{a\mu}\right) \to \left(X^\mu, P_\mu, x^\mu, p_\mu\right)$
\begin{gather*}
x=x_1-x_2, \qquad   P=p_1+p_2, \qquad M=\sqrt{P^2},
\nn\\
p=p_1-\xi(M)P, \qquad X=\xi x_1+(1-\xi)x_2 +
\frac{d\xi}{dM}(P\cdot x)\hat P,\qquad \hat P=P/M, \nn
\end{gather*}
where $\xi(M)$ is an arbitrary function, and introduce the
operators
\begin{gather*}
\mcl{H}^{\rm free} = \gamma_2^\pa(\gamma_1\cdot p_1-m_1) +
\gamma_1^\pa(\gamma_2\cdot p_2-m_2)
\nn\\
\phantom{\mcl{H}^{\rm free}}{}=\gamma_1^\pa\gamma_2^\pa M +
\gamma_2^\pa(\gamma_1\cdot p-m_1) +
\gamma_1^\pa(-\gamma_2\cdot p-m_2), \nn \\
\mcl{K}=\tfrac{1}{2}(\gamma_1\cdot p_1+m_1)(\gamma_1\cdot p_1-m_1)
- \tfrac{1}{2}(\gamma_2\cdot p_2+m_2)(\gamma_2\cdot p_2-m_2)
\nn\\
\phantom{\mcl{K}}{} =\tfrac{1}{2}(p_1^2-p_2^2-m_1^2+m_2^2)=P\cdot
p - \nu(M), \nn
\end{gather*}
such that $[\mcl{H}^{\rm free},\mcl{K}]  = 0$; here
$\gamma_a^\pa\equiv\gamma_a\cdot\hat P$ and $\nu(M)=
\tfrac{1}{2}\left(m_1^2-m_2^2+(1-\xi(M))M^2\right)$. In these
terms the equations~\re{A.1} take the equivalent form
$\mcl{H}^{\rm free}\iPhi=0$ and $\mcl{K}\iPhi=0$.

    At this point we choose the mixed representation for the wave
function: $\iPhi=\iPhi(P,x)$. Then it follows from the constraint
$\mcl{K}\iPhi=0$
\[
\iPhi(P,x)=\exp\left(\im\frac{\nu(M)}{M}x^\pa\right)
\tilde\iPhi(P,x_{\ort}),
\]
where $x^\pa=\hat P\cdot x\stackrel{\B P=0}{\longrightarrow}x^0$
and $x^\mu_\ort=(\delta^\mu_\nu-\hat P^\mu\hat
P_\nu)x^\nu\stackrel{\B P=0}{\longrightarrow}(0,\B r)$ thus
$\tilde\iPhi(P,x_\ort)\stackrel{\B
P=0}{\longrightarrow}\tilde\iPhi(E,\B r)$ where
$E=\left.P_0\right|_{\B P=0}=\left.M\right|_{\B P=0}$. In other
words, this constraint suppresses the $x^\pa$-dependence of
$\tilde\iPhi$ and eliminates the relative time $x^0$ from the
center-of-mass description. Equivalently, one can choose
$\nu(M)=0$, then $\iPhi$ is free of $x^\pa$. We note that the
choice of $\nu(M)$ or $\xi(M)$ does not af\/fect the form of the
operator $\mcl{H}^{\rm free}$ and so it is gauge-f\/ixing. Thus we
do not distinguish the functions~$\iPhi$ and $\tilde\iPhi$ any
more.

    In the center-of-mass (c.m.) frame of reference (where $\B P=0$) we have
\[
\mcl{H}^{\rm free}\iPhi(P,x_{\ort}) \stackrel{\B
P=0}{\longrightarrow} \beta_1\beta_2\left\{E-h_1(\B p)-h_2(-\B
p)\right\}\iPhi(\B r)=0,
\]
where $\B p=\B p_1=-\B p_2$,
$\beta_a\equiv\gamma^0_a=\left.\gamma^\pa\right|_{\B P=0}$, and
operators $h_a(\B p)$ are def\/ined in \re{2.2}. Up to the factor
$-\beta_1\beta_2$ this equation coincides with 2BDE \re{2.1} in
the case of free particles.

    In the general case the interaction term $\mcl{V}\iPhi$
appears in the r.h.s.\ of the equation $\mcl{H}^{\rm free}\iPhi=0$
(instead of zero), and we have the following set of equations
\[
P\cdot p\,\Psi=0 \qquad \mbox{and} \qquad \mcl{H}^{\rm
free}\Psi=\mcl{V}\Psi.
\]
Since a free-particle operator $\mcl{H}^{\rm free}$ is
Poincar\'e-invariant the potential $\mcl{V}$ must be so. Besides,
it must obey the equalities
\begin{gather}\lab{A.8}
\mcl{V}_{\rm c.m.}\equiv\left.\mcl{V}\right|_{\B
P=0}=\beta_1\beta_2U(\B r)
\end{gather}
and $[\mcl{V},P\cdot p]=0$. These requirements allow us to
construct unambiguously the covariant opera\-tor~$\mcl{V}$ for
arbitrary potential of the form \re{2.3}.

    It follows from \re{A.8}, \re{2.3} that the operator
$\mcl{V}_{\rm c.m.}$ consists of the sum of partial poten\-tials~$U_A(r)$ multiplied by matrix coef\/f\/icients
$\beta_1\beta_2\Gamma_A$. Both factors in every term of the sum
are $O(3)$-invariant. We construct Lorentz-scalar counterparts for
them separately.

    Let $\rho=\sqrt{{-}\vphantom{x^2}\smash{x^2_\ort}}$ and thus
$\rho_{\rm c.m.}=|\B r|=r$. Then functions $U_A(\rho)$ are Lorentz
scalars by construction and $\left.U_A(\rho)\right._{\rm
c.m.}=U_A(r)$.

 \re{2.5}--\re{2.6} is a basis for matrices $\beta_1\beta_2\Gamma_A$:
$\beta_1\beta_2\Gamma_{\rm ee}=\Gamma_{\rm ee}$,
$\beta_1\beta_2\Gamma_{\rm oo}=\Gamma_{\rm oo}$,
$\beta_1\beta_2\Gamma_{\rm eo}=\im\Gamma_{\rm eo}$,
$\beta_1\beta_2\Gamma_{\rm oe}=\im\Gamma_{\rm oe}$ (the factor
``$\im$'' is not important in this regard). First of all it is
convenient to express $\B\sigma$-matrices in the spin factors
\re{2.7}, \re{2.8} of the basis \re{2.5}--\re{2.6} via
$\B\gamma$-matrices as follows:
$\B\sigma_a=\gamma_a^5\beta_a\B\gamma_a$. Then every basis element
is a product of factors
\[
I,\quad \gamma_a^5,\quad \beta_a,\quad
\B\gamma_1\cdot\B\gamma_2,\quad\B\gamma_a\cdot\B n, \quad (\B
n,\B\gamma_1,\B\gamma_2)
\]
which are scalar or pseudo-scalar with respect to the group
$O(3)$. We construct the Lorentz scalar or pseudo-scalar
counterparts of these matrices as follows
\begin{gather}
I\ \dashrightarrow\ I, \qquad   \gamma_a^5\ \dashrightarrow\
\gamma_a^5, \qquad  \beta_a\ \dashrightarrow\ \gamma_a^\pa,\qquad
\B\gamma_1\cdot\B\gamma_2\ \dashrightarrow\
-\gamma_1^\ort\!\cdot\gamma_2,
\nn\\
\B\gamma_a\cdot\B n\ \dashrightarrow\ -\gamma_a\cdot n_\ort,\qquad
(\B n,\B\gamma_1,\B\gamma_2)\ \dashrightarrow\ -(\hat
P,n,\gamma_1,\gamma_2)\equiv
-\varepsilon_{\mu\nu\lambda\varkappa}\hat P^\mu
n^\nu\gamma_1^\lambda\gamma_2^\varkappa,\lab{A.9}
\end{gather}
where $n=x/\rho$ and $\varepsilon_{\mu\nu\lambda\varkappa}$ is
absolutely antisymmetric pseudo-tensor ($\varepsilon_{0123}=-1$).

    The resulting covariant potential $\mcl{V}$ has the form
\[
\mcl{V}(x,\hat P)=\sum\nolimits_{A=1}^{48} U_A(\rho)\Xi_A,
\]
where matrices $\Xi_A$ are built with $\beta_1\beta_2\Gamma_A$ by
the replacement~\re{A.9} (with an unchanged order of co-factors).

    The unambiguity of this construction is obvious: if there exist two
potentials which coincide in the c.m. reference frame then their
dif\/ference is zero in any reference frame, due to
Poincar\'e-invariance of these potentials.

\pdfbookmark[1]{References}{ref}
\LastPageEnding

\end{document}